\documentclass{vgtc}                          

\ifpdf
  \pdfoutput=1\relax                   
  \usepackage{graphicx}                
  \DeclareGraphicsExtensions{.pdf,.png,.jpg,.jpeg} 
\else
  \usepackage{graphicx}                
  \DeclareGraphicsExtensions{.eps}     
\fi%

\graphicspath{{figures/}{pictures/}{images/}{./}} 

\usepackage{microtype}                 
\PassOptionsToPackage{warn}{textcomp}  
\usepackage{textcomp}                  
\usepackage{mathptmx}                  
\usepackage{times}                     
\usepackage{cite}                      
\usepackage{tabu}                      
\usepackage{wrapfig}
\usepackage{booktabs}   
\usepackage{booktabs}
\usepackage{soul}
\usepackage{color}
\usepackage{amssymb}
\usepackage{amsmath}

\usepackage{algorithm,algorithmic}
\usepackage{subfigure}

\onlineid{5799}

\vgtccategory{Research}

\vgtcinsertpkg

\newcommand{\name}{PromptAid}

\title{\name: Prompt Exploration, Perturbation, Testing and Iteration using Visual Analytics for Large Language Models}

\author{Aditi Mishra\thanks{e-mail: amishr45@asu.edu} \\ \scriptsize Arizona State University %
\and Utkarsh Soni\thanks{e-mail:usoni1@asu.edu} \\ \scriptsize Arizona State University
\and Anjana Arunkumar\thanks{e-mail:aarunku5@asu.edu} \\ \scriptsize Arizona State University
\and Jinbin Huang\thanks{e-mail:jhuan196@asu.edu}\\ \scriptsize Arizona State University
\and Bum Chul Kwon\thanks{e-mail:bumchul.kwon@us.ibm.com} \\ \scriptsize IBM Research
\and Chris Bryan\thanks{e-mail:cbryan16@asu.edu}\\ \scriptsize Arizona State University}

\teaser{
  \centering
  \includegraphics[width=\linewidth]{figures/Interface.pdf}
  \caption{The \textsc{\name} interface, shown here with the ag\textunderscore news dataset, consists of six main linked sections which support \textsc{(A)} selecting models, domains and entering custom prompt templates, \textsc{(B)} exploring the prompt space, \textsc{(C)} analyzing instance level performance of a prompt template, \textsc{(D)} comparing versions of prompt templates over multiple iterations, \textsc{(E)} obtaining recommendations for prompt template alteration, and \textsc{(F)} testing generated templates on in or out of distribution data points.
  }
  \label{fig:teaser}
}

\abstract{Large Language Models (LLMs) have gained widespread popularity due to their ability to perform ad-hoc Natural Language Processing (NLP) tasks with a simple natural language prompt. Part of the appeal for LLMs is their approachability to the general public, including individuals with no prior technical experience in NLP techniques. However, natural language prompts can vary significantly in terms of their linguistic structure, context, and other semantics. Modifying one or more of these aspects can result in significant differences in task performance. Non-expert users may find it challenging to identify the changes needed to improve a prompt, especially when they lack domain-specific knowledge and lack appropriate feedback. To address this challenge, we present \textsc{\name}, a visual analytics system designed to interactively create, refine, and test prompts through exploration, perturbation, testing, and iteration. \textsc{\name} uses multiple, coordinated visualizations which allow users to improve prompts by using the three strategies: keyword perturbations, paraphrasing perturbations, and obtaining the best set of in-context few-shot examples. \textsc{\name} was designed through an iterative prototyping process involving NLP experts and was evaluated through quantitative and qualitative assessments for LLMs. Our findings indicate that \textsc{\name} helps users to iterate over prompt template alterations with less cognitive overhead, generate diverse prompts with help of recommendations, and analyze the performance of the generated prompts while surpassing existing state-of-the-art prompting interfaces in performance.
} 

\CCScatlist{
  \CCScatTwelve{Human-centered computing}{Visu\-al\-iza\-tion}{Visu\-al\-iza\-tion techniques}{Treemaps};
  \CCScatTwelve{Human-centered computing}{Visu\-al\-iza\-tion}{Visualization design and evaluation methods}{}
}

\begin{document}


\maketitle
\section{Introduction}

\maketitle

Large Language Models (LLMs) are neural networks trained on a large corpus of unlabelled data in a self-supervised manner. They have brought about a paradigm shift in the field of Natural Language Processing (NLP)~\cite{liu2023pre}. In addition to achieving state-of-the-art performance across various NLP tasks~\cite{scao2021many}, such as translation, named entity recognition, and question answering, LLMs are widely accessible to non-technical users because they can easily adapt the models for specific downstream tasks using plain text (i.e., human-like language)~\cite{liu2023pre}. 
This process of giving natural language instructions to an LLM is called \textbf{prompting}; in part, the ease and intuitiveness of prompting has led to the widespread adoption of tools such as ChatGPT~\cite{million}.  
A prompt typically consists of task instruction with optional input and expected output examples, followed by the task itself. 
For example, one could prompt an LLM with the instruction: \textit{Determine the sentiment of the following review. An example of a review is ``The book was a fun read'' and the sentiment is `positive'}. 
The LLM can then be prompted using the same template to determine the sentiment of a new review, such as \textit{The book was long and boring}. 

Prompting enables treating LLMs as few shot learners~\cite{brown2020language}, where a trained LLM can perform downstream tasks such as classification, summarization, question answering, and so on, solely based on giving it a small number of (``a few'') instructions and task demonstrations via prompts. 
The LLM can then learn the downstream task without needing to re-train on a new training dataset or update the parameters of the underlying model. 
Recent work has highlighted the success of LLMs as few-shot learners, wherein they show comparable performances to traditional AI/ML approaches~\cite{radford2019language}.


Despite the great promise and remarkable progress of LLMs, it is yet difficult for human users to craft optimal prompts and priming examples that generate desired outputs consistently~\cite{jiang2022promptmaker,zamfirescu2023johnny}. 
As a result, several prompting strategies have been proposed, including identifying prompts that lead to more errors, identifying why those errors occur, and finally resolving the error to bias the language model correctly~\cite{zamfirescu2023johnny}. 
A significant challenge is that, while LLMs are accessible to non-technical users via natural language prompting, adapting the language model appropriately requires domain knowledge of the downstream task and multiple prompt iterations of creating, refining, and analyzing prompts.
This challenge gave rise to a new type of NLP job title called \textit{prompt engineer} ~\cite{liu2023pre}. 
It also raises the question of how a non-expert user who does not have expertise in NLP can create and improve the performance of LLMs for target tasks using prompt engineering.
Many prior studies stress the importance of user-friendly interfaces to aid prompting for non-expert users~\cite{jiang2022promptmaker,liu2021makes,zamfirescu2023johnny}.

To address the challenges, we present a visual analytics system called \textsc{\name{}} for non-expert users to explore, perturb, refine, and test prompt templates iteratively.
We designed \textsc{\name{}} based on a pre-study with three NLP experts.
The system consists of multiple, coordinated visualizations for exploring prompts and supports three different semi-automated strategies for improving prompts: keyword suggestions, paraphrases, and in-context example recommendations.
To demonstrate the usefulness of \textsc{\name{}}, we conducted two case studies. We also conducted a within-subject experiment where non-expert users were asked to prompt LLMs so that they can successfully classify topics of text input by using \name{} and a baseline tool each.
Our study results show that \textsc{\name{}}'s features such as interactive visualizations and recommendations for prompt perturbation and iteration were regarded as highly useful for the tasks by the participants.

The main contributions of this paper include the following: (1)~We analyze design challenges and goals for prompting by non-experts, with a focus on optimizing performance with reduced cognitive overhead, based on a pre-study with NLP researchers and by reviewing recent literature. (2) We develop \textsc{\name{}}, a visual analytics interface that lets a user interactively and semi-automatically perturb prompts both \textit{linguistically} and \textit{contextually}, based on system recommendations, to obtain better accuracies using open sourced language models. (3) Based on our experience in creating and robustly evaluating \textsc{\name{}}, we discuss how interactive visualization-driven prompt crafting can increase user efficiency while lessening cognitive effort.

\vspace{-2mm}

\begin{figure*}[t]
  \centering
  \includegraphics[width=1\textwidth]{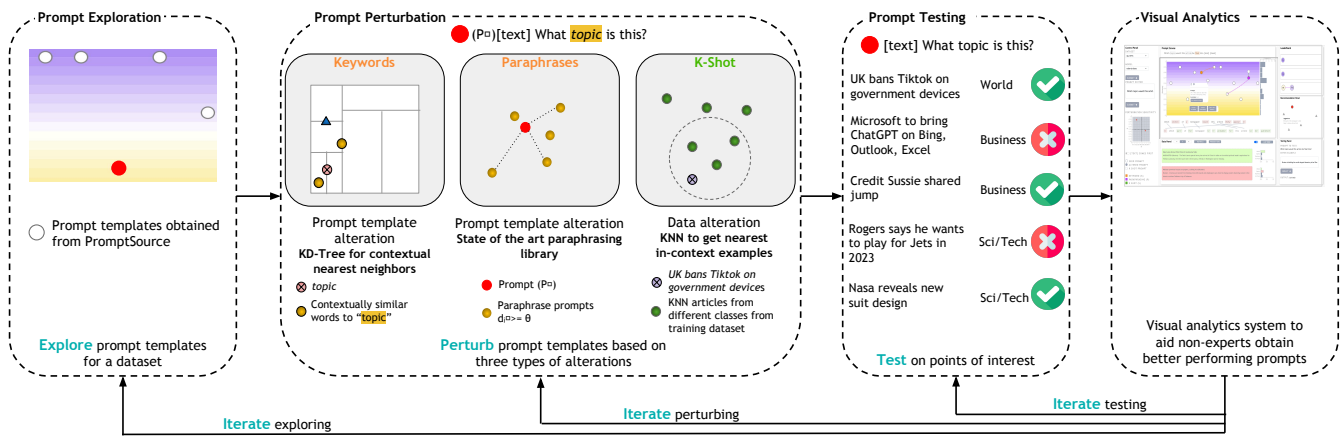}
  \vspace{-3mm}
  \caption{\textsc{\name} employs a multi-phase approach: templates are embedded in a latent space and clustered based on similarity in the exploration phase. In the perturbation phase, contextual keywords, paraphrases, and in-context examples are recommended using KD-Tree, the Parrot library, and KNN, respectively. Users can then test alterations on data points of interest in the testing phase. The frontend interface employs visual analytics to streamline these processes iteratively, leading to the generation of desired prompt templates.}

  \label{fig:workflow}
  \vspace{-4mm}
\end{figure*}

\section{Related Work}
\label{sec:related_work}
In this section, we first review prior literature on large language models and prompt engineering. Then, we identify a research gap in visual analytics to support prompt engineering for few-shot learning LLMs.

\subsection{Language Models and Prompt Engineering}
While the emergence of LLMs is relatively recent, the first language model was developed almost 60 years ago; the first model, named ELIZA, used pattern recognition and a rules-based logic to mimic human conversation~\cite{weizenbaum1966eliza}.
The recent breakthrough of AI research called transformers \cite{vaswani2017attention} that are used to model sequential data (e.g., natural language) by using the self-attention approach,  along with the vast amount of data available on the internet, led to the development of LLMs, which include BERT \cite{DBLP:journals/corr/abs-1907-11692}, RoBERTa \cite{DBLP:journals/corr/abs-1907-11692}, ALBERT \cite{DBLP:journals/corr/abs-1909-11942} and GPT-2,3~\cite{radford2019language} and more recently GPT-4~\cite{gpt4}. 
Broadly, LLMs are believed to capture the semantics and syntax of human language because they are trained with large parameters and large datasets. 
The embeddings from these pre-trained models can then be fine-tuned using a small dataset to perform more narrowly defined, downstream NLP tasks including text classification, summarization, knowledge retrieval, and so on. 
This can be done by adding task-specific layers to the end of LLMs and updating partial or all parameters with the backpropagation method. 
This paradigm to solve a specific task is called pre-train + fine-tune~\cite{radford2018improving,sarzynska2021detecting,yang2019xlnet}.

However, in the past two years, NLP has seen a paradigm shift from pre-train + fine-tune to pre-train + prompt~\cite{liu2023pre}. 
Given a task description or context in natural language, the LLM can be \textit{prompted} to result in an output for downstream tasks without requiring changes to the underlying model. 
Brown et al.~\cite{brown2020language} demonstrated that GPT-3 can handle a wide variety of NLP tasks with only a few task demonstrations and prompts as context. 
In particular, the idea of prompting has made large ML models significantly more accessible to non-expert users and is increasingly being used in various applications~\cite{liu2023pre}.

However, to elicit appropriate answers from an LLM, a prompt must be properly engineered~\cite{liu2023pre}. 
As human language can be highly nuanced and varied, writing prompts in different ways (even with subtle changes) can lead to different responses from the language model. 
Recent studies~\cite{jiang2022promptmaker,zamfirescu2023johnny} identified several pain points faced by non-expert users in prompting. 
According to prior research \cite{jiang2022promptmaker,zamfirescu2023johnny,strobelt2022interactive}, prompting challenges can be classified into either of the two categories: linguistic or contextual:
1) \textbf{Linguistic}: There are many ways that a desired prompt can be formulated, and altering components such as word choice, phrasing, length, prefixes, and other linguistic elements can significantly impact the accuracy of the task, even holding consistent the prompt's other components. 
As a result, current solutions employ a brute force strategy to generate various prompt combinations~\cite{strobelt2022interactive}.
2) \textbf{Contextual}: In a few-shot setting, developing an optimal set of priming examples requires a user to select a small set of examples that effectively represent the desired task and can achieve high accuracy. 
For a non-expert user (e.g., someone without sufficient expertise to identify the most effective examples for a given task, or set the number of $k$ examples which might optimize a prompt's performance), this can be especially challenging and cause significant cognitive load~\cite{jiang2022promptmaker}.

To alleviate the problems with prompt engineering, Mishra et al. ~\cite{mishra2022help} introduced a technique in which an LLM is used to generate task-specific questions which the user can answer. 
This was shown to provide better context for the LLM for a downstream task.
However, there is not yet a solution that can automatically or semi-automatically find the best prompts for desired tasks with given LLMs without undergoing iterative human-in-the-loop processes.

\subsection{Visual Analytics for Prompt Engineering and NLP}

Due to the newness of the pre-train + prompt paradigm, there has been little work thus far in the HCI and Visualization research communities to develop techniques and/or interfaces to augment the prompting process. The closest work to this paper is PromptIDE~\cite{strobelt2022interactive}, which provides users with an interface to experiment with prompt variations, visualize their performance, and subsequently try to optimize the prompts. 
However, in contrast to PromptIDE, \textsc{\name{}} (i) generates prompt recommendations that can be refined by users and (ii) supports zero- and few-shot learning.
For general visual analysis of NLP models, several prior tools can broadly be categorized as supporting two types of tasks: (1)~greying black boxed NLP models \cite{strobelt2017lstmvis,hoover2019exbert,jaunet2021visqa,wang2021dodrio,derose2020attention,vig2020bertology,vig2019multiscale,vig2019analyzing}, primarily for NLP model developers and experts, and (2)~understanding post hoc model behavior based on input variances~\cite{tenney2020language, liu2018nlize}, 

One relevant tool is the Language Interpretability Tool (LIT)~\cite{tenney2020language}, which supports analysis of NLP models at the global and local levels by visualizing embedding spaces, saliency, accuracy metrics, and more. We were in part inspired by this tool to treat each prompt template in the work as an ``NN model,'' supporting both global and instance-level analysis, along with projecting the prompts into an embedding space (see Section~\ref{sec:design_challenges}). Moreover, we support users validating their understanding of generated prompts by providing them the ability to test their ``models.'' However, in contrast to the LIT, \textsc{\name{}} supports an iterative prototyping approach based on semi-automated recommendations and human feedback (and also focuses on prompt engineering, not general NLP models). 

Another relevant tool is NLIZE~\cite{liu2018nlize}, which lets users visually perturb an natural language inference model's internal hidden states to evaluate model outputs. We expand upon this perturbation approach: NLIZE does not let users perturb the input being sent to the model, but \textsc{\name{}} lets novice users alter the input provided to the LLM based on system recommendations.



\section{Design Challenges and Goals}
\label{sec:design_challenges}

In line with the prior visualization interface targeted for prompting~\cite{strobelt2022interactive}, we identify a set of design challenges (C1-C3) that novice users encounter when interrogating language models, based on a pre-study with three NLP experts.
We chose NLP experts who were familiar with prompting LLMs for tasks without requiring re-training and fine-tuning. 
Based on this pre-study, we derived the five key design goals (G1-G5) that guided the design of \textsc{\name{}}. 

\subsection{Design Challenges}

Each of the design challenges corresponds to a specific process that \textsc{\name{}} endeavors to facilitate for novice users. 
These processes are denoted by labels presented in blue-gray boxes.

\textbf{(C1)}{\textsc{Exploration}}; \textbf{Navigating and exploring a vast prompt space is difficult.}
Constructing natural language prompts requires synthesizing intricate linguistic components such as keywords, phrasing, and structure, all of which can impact the output of an LLM~\cite{jiang2022promptmaker}. The space of possible natural language prompts can also rapidly expand, leading to a multitude of performances that can be challenging for non-expert users to navigate when they want to identify an optimal prompt template. 
Current brute force solutions can impose a high cognitive load on users~\cite{strobelt2022interactive}, which can be further complicated by the tendency of humans to overgeneralize from single failures~\cite{zamfirescu2023johnny,yang2018grounding}. 
In other words, a prompt might perform well globally but poorly on a specific instance, and users might underestimate its overall effectiveness. \textit{\textbf{There is a need for systems that allow users to explore the prompt space and analyze the performances of prompt templates}}. 
Such insights can ultimately lead users to craft optimal prompts for their tasks.

\textbf{(C2)} {\textsc{Perturbation}} \textbf{A high cognitive effort is required to source words, paraphrase prompts, and obtain priming examples.} 
Though LLMs demonstrated impressive capabilities in  generalizing to new tasks with only a few examples, previous studies~\cite{jiang2022promptmaker,zamfirescu2023johnny,mishra2022help} revealed that generating these changes and selecting the k-shot examples necessary for these new tasks entail higher cognitive effort for non-expert users. 
Furthermore, the k-shot examples chosen by users can significantly influence the results on new data, which may result in under- or over-generalization of the prompts~\cite{jiang2022promptmaker}. 
With vast amounts of data available on the Internet, users may attempt to find similar words for instruction and come up with k-shot examples for testing, but there is no guarantee that the words chosen for instruction capture users' intentions accurately and that the k-shot examples are optimal for their intended purpose. 
Even when employing AI-based recommendations, human intervention is required to ensure that the task's semantics remain unchanged during the perturbations. 
Consequently, \textit{\textbf{interactive systems can be instrumental in assisting users in identifying suitable suggestions for instruction and generating examples}}.

\textbf{(C3)} {\textsc{Testing}} \textbf{Evaluating prompts in global and instance levels is challenging.} Finally, LLMs are inherently stochastic (as they use a statistical model to predict the probability of the next word), meaning that the answers produced by the model can vary for the same prompt. Previous studies (e.g.,~\cite{jiang2022promptmaker}) have identified this as a significant challenge for prompt evaluation, as users need more global metrics to comprehend how a prompt performs over a small test set, instead of relying on a single data point. They also highlight the user's needs to compare the performance of diverse prompts on a small representative test set.
However, presenting such results in a raw and tabular format can be cognitively taxing for non-expert users; this makes visualization a promising approach, due to its ability to graphically encode complex and abstract information into meaningful representations. \textit{\textbf{Therefore, visualizations need to provide a view for evaluation of prompt performances, accompanied by global and instance metrics for users to compare the performance of prompts. Furthermore, users should be able to evaluate a prompt based on their custom-curated examples.}}


\subsection{Design Goals}

The following design goals were derived with help of NLP experts to address the design challenges described above.
We provide the corresponding challenges wherever appropriate.

\textbf{(G1) Provide an overview of prompts retrieved or altered.} The increasing adoption of language and image-based prompting models (e.g., Dall.E) has resulted in a rapid expansion of datasets containing crowd-sourced prompts tailored to specific tasks, such as summarizing lengthy text paragraphs or facilitating natural language question-answering~\cite{bach2022promptsource,wang2022diffusiondb,mishra2021cross}. In preliminary studies, multiple participants observed that enabling end-users to navigate, explore, and analyze prompts based on their semantic similarity and performance \textbf{(C1)} could aid in the design of better prompts. \textit{\textbf{As such, the application of visualization techniques can offer a means to present users with an overview of prompts, grouping them according to similarity and performance.}}

\textbf{(G2) Support both global and instance-level analysis of prompts.} 
Testing on a single data point might not give users sufficient insight into how well (or poorly) each prompt performs. 
The stochastic nature of LLMs must also be accounted for. 
Thus, users should be able to test prompts and obtain a heuristic accuracy on smaller datasets in a faster feedback loop for prototyping purposes \textbf{(C1, C3)}. 
Further, only explaining a prompt from a global perspective might lead the user to overlook important details, resulting in misinterpretations of why a prompt performed better or worse. 
To enable detailed inspection, while not overwhelming users by revealing too many details,  \textit{\textbf{visualizations need to show data points (examples), their predicted probabilities over classes, and other accuracy metrics such as precision and recall to help analyze and compare the performance of multiple prompts.}}

\textbf{(G3) Provide recommendations for prompts.} While G1, G2 are crucial for the comprehension of the overall behavior of a set of prompts based on their semantic similarities, global, and instance-based performance, however these metrics do not offer users the means to modify the prompts to improve results. A prompt's efficacy can be influenced by linguistic elements, such as the selection of keywords and sentence phrasing, as well as contextual components, such as the use k-shot examples. As such, the process of varying not only the linguistic components but also the selection of the best examples for the k-shot setting can be mentally taxing for the end user. To address this issue, \textit{\textbf{the interface should support both linguistic and contextual modifications, and provide visualizations that recommend prompt changes in a way that not only eases cognitive overload but also aids end users in creatively thinking in new directions}} \textbf{(C2).}

\textbf{(G4) Provide visual steering and immediate feedback for changed prompts.} 
While running preliminary experiments for this work, we realized that the prompt space could expand exponentially. 
This is because there could be numerous changes that can be made to the original prompt while maintaining the semantics. 
The prompt space can further expands in size if we allow for contextual changes as well.
It is challenging to read the individual prompts in the text without visual aid.
\textit{\textbf{Thus, visual steering can make navigating the prompt space less cognitively demanding}} \textbf{(C2)}. 

\textbf{(G5) Allow users to generate custom examples.} While the heuristic accuracy of a prompt template can be calculated, users might also want to enter examples of their choices and observe the LLM's outputs. This process can aid in evaluating the robustness of the prompt template on various examples \textbf{(C3)}. 
\textit{\textbf{Thus, the interface should provide users the ability to test custom-curated samples.}}


\section{Backend System Design}
In this section, we discuss the design of the \textsc{\name{}} system by describing how different views support the design goals \textbf{(G1-G5)}. 
Figure \ref{fig:workflow} shows an overview of the system pipeline which consists of a backend module (this Section) and a frontend interface (described in Section~\ref{sec:frontend}). 
The pipeline integrates several methods and techniques, such as K-D Trees, K-NN, a state-of-the-art paraphrasing framework, and LLMs to automatically extract relevant recommendations based on the changes the user wishes to make on a given prompt template.

\subsection{Dataset and Model}
\label{sec:dataset_and_model}
To showcase the feasibility of \textsc{\name{}} for NLP tasks, we support two common tasks, namely topic classification and sentiment analysis, using the benchmark datasets of ag$\textunderscore$news~\cite{Zhang2015CharacterlevelCN} and amazon$\textunderscore$polarity~\cite{mcauley2013hidden}, respectively. 
We utilize a set of prefix seed prompt templates~\cite{liu2023pre} retrieved from the PromptSource library~\cite{bach2022promptsource}.

To test the prompt templates, we integrate three open-source language models into \textsc{\name{}}: RoBERTa-base, GPT-2, and T0pp, implemented using the OpenPrompt library~\cite{ding2021openprompt} in Python. 
The results are calculated based on user interactions on the fly using a Node.js server.
While we test a prompt template on a modest sample of 20 data points, it is worth noting that \textsc{\name{}} is task-independent and model-agnostic, so it can be easily adapted to other NLP tasks and models; see Appendix~\ref{sec:usage_of_smaller_testing_set} for a discussion of why we opt with for a testing set of this number of data points, including scalability and user experience considerations.

\subsection{Keyword, Paraphrasing, and K-Shot recommendations}
\label{sec:key_para_kshot}
Building upon prior work that has explored the challenges faced by non-expert users in generating effective prompts for language models \cite{jiang2022promptmaker,zamfirescu2023johnny}, we consider the impact of keyword and phrasing choices (linguistic challenges), as well as the use of k-shot examples (contextual challenges) to inform prompt crafting. While factors such as prompt length, structure, and k-shot ordering may also affect prompt performance, these require further investigation --- see Section~\ref{sec:discussion}.

In our system, we allow users to choose linguistic alterations from the recommendation panel only, as these directly affect the prompt template. K-shot examples are dynamically calculated and presented to the user, as they pertain to each specific data point, and do not alter the prompt template. While our initial prototypes considered allowing users to choose their own k-shot examples, we found that this approach could be overwhelming for a novice user, as it would require the selection of examples for each data point in the test set to obtain an accurate global accuracy metric. To reduce cognitive load, we instead provide users with pre-selected recommendations for k-shot examples. That said users can enter their few-shot examples and prompt templates in the Prompt Editor panel to obtain predictions. 

When providing linguistic recommendations, we consider the trade-off between \textit{relevancy} and \textit{diversity}. We define relevancy as the degree to which recommended words or phrases are closely related to the prompt that is being modified, while diversity refers to recommendations that are more distant in the embedding space. In some domains, such as luxury fashion, diversity of recommendations may be more important than relevancy, as rare and exclusive items are highly valued~\cite{sa2022diversity}. Conversely, in e-commerce applications, relevant recommendations that align with specific user goals are more valuable. To our knowledge, prior research has not yet investigated the effects of altering recommendations towards more diverse words; we currently assign equal weights to both relevancy and diversity in our recommendation output and have found this works well (e.g., see use cases and user study in Sections~\ref{sec:case_studies}, \ref{sec:evaluation} and \ref{sec:usecase_2} in the Appendix). However, further research and testing are needed to optimize these trade-offs; \textsc{\name{}}'s weights can be easily updated based on future findings.

\textbf{Keyword Recommendations.}
Prior research has demonstrated that the choice of keywords in a prompt template can significantly impact its effectiveness \cite{jiang2022promptmaker}. Specifically, prompts containing words that explicitly specify the task at hand can outperform those with abstract descriptions of the same task \cite{reynolds2021prompt}. To facilitate the creation of effective prompt templates, we use K-D trees to identify words that are contextually similar to the word the user intends to change, for the specific task being performed.

To accomplish this, we first append the task type to the end of the prompt template $P_t(.)$, and send the resulting prompt to sentence transformer models (LLMs) to obtain contextual embeddings of the words. We use a K-D tree to identify words that are similar to the word the user intends to alter. 
To do this, we obtain the nearest words from a 10,000-word public web corpus~\cite{goldhahn2012building}, which is vectorized using the same sentence transformer model. From this set, we select twenty nearest words and then choose five words closest and five words farthest from the word being altered. 
To maintain semantic coherence in the prompt template after substitution and to avoid repetitions, we additionally perform lemmatization to remove words with the same root word.

\textbf{Paraphrase Recommendations.} Paraphrasing-based approaches aim to generate candidate prompt instructions that are similar to the seed prompt while being sufficiently different to offer a range of options. 
An ideal paraphrase should preserve the meaning of the original prompt instruction, be grammatically correct, and differ from the seed prompt~\cite{prithivida2021parrot}. To achieve this, we employ a state-of-the-art paraphrasing library called Parrot~\cite{prithivida2021parrot}. Notably, this library supports parameters to account for both relevancy and diversity in paraphrased recommendations. However, in practice, we found that in certain cases, the library returned paraphrased statements that were highly similar to the seed prompt template $(P_t(.))$ (e.g., only changing a single word). To ensure that the paraphrased prompt template maintains sufficient distinction, we define a threshold $\theta$ based on pairwise Levenshtein distances between the seed prompt and the paraphrases, as well as between the paraphrases themselves, and exclude any new paraphrases with a distance $\leq \theta$.

To determine the appropriate threshold, we conducted a set of preliminary experiments and found that for seed prompt templates with length $<10$, a threshold of $\theta=20$ tended to produce paraphrases sufficiently different from the seed prompt. For seed prompt templates longer than 10 characters, we set a threshold of $\theta=25$. In practice, we found these values tended to balance between relevant and diverse paraphrases that were both meaningful and distinct from the seed prompt. 

\textbf{K-Shot Example Recommendations.} Prior research has underscored the importance of example selection in few-shot learning settings, with the choice of examples potentially resulting in outcomes ranging from state-of-the-art performance to random guessing \cite{lu2021fantastically}. To address this issue, prior studies have attempted to identify an optimal set of examples that can yield the best results from a language model \cite{liu2021makes,gao2020making}. Building on this approach, we propose a K-NN-based approach to identify an optimal set of k-shot examples.

To accomplish this, we first identify the five nearest examples $x_1, x_2, \dots, x_5$ to a given test point $x_{test}$ from the training dataset $D_{train}$, which in our case were the a$\textunderscore$news and the amazon$\textunderscore$polarity datasets. We limit the number of examples to five, as we found that adding additional examples (for both task classification and sentiment analysis) did not significantly improve the performance of the prompt template. Similarly, Lie et al.~\cite{liu2021makes} found that increasing the number of examples did not lead to improved performance on an IMDB dataset for a sentiment analysis task. This also helped provide a more responsive user experience, as increasing $k$ led to longer processing times.

To identify the k-shot examples, we convert both the test data point $x_{test}$ and the data points in the training dataset $D_{train}$ into vector representations using a general-purpose sentence transformer model (specifically, the \textit{all-MiniLM-L6-v2} model~\cite{reimers-2019-sentence-bert}. Then, for each test point, we identify the five nearest neighbors from the $D_{train}$. We found that the nearest neighbors of the test point from the train set generally belonged to the same class as the test point, which could lead to the language model merely copying the label, resulting in a bias towards the example point labels. To mitigate this issue, if $k>1$, we chose the $(k-1)$ closest neighbors from classes different from that of the test data point and one which belonged to the same class as the test data point, else we choose the nearest point from another class to the test data point.

Finally, we order the k-shot examples based on their cosine distances using the following inequality: $d(x_{test},x_i) < d(x_{test},x_j)$, where $x_i, x_j\epsilon D_{train}$.
These examples are inserted into the prompt template $P_t(x_i)$ being tested and concatenated $[P_t(x_1)y_1,P_t(x_2)y_2,\dots P_t(x_k)y_k]$ to form the k-shot example set.
This example set is then sent along with $x_{test}$ to the LLM. To identify the optimal value of $k$ in the k-shot setting, we iteratively ran the concatenated examples with $k\epsilon[1,5]$ and returned the $k$ with the highest accuracy metrics.

\subsection{Perturbation Sensitivities}
``Next step'' linguistic perturbation sensitivities in \textsc{\name{}} are calculated to visually steer non-expdert users towards the ``right'' perturbation choice (i.e., keywords or paraphrasing) in the next step, to ultimately increase the performance of the prompt template. We obtain these sensitivities through the sampling of prompt templates for each perturbation type from the initial template and subsequently measuring their average heuristic accuracy on the test dataset. Refer to section \ref{sec:key_para_kshot} to see why we choose to showcase only linguistic perturbations.

\begin{figure*}[t]
   \centering
    \includegraphics[width=\linewidth]{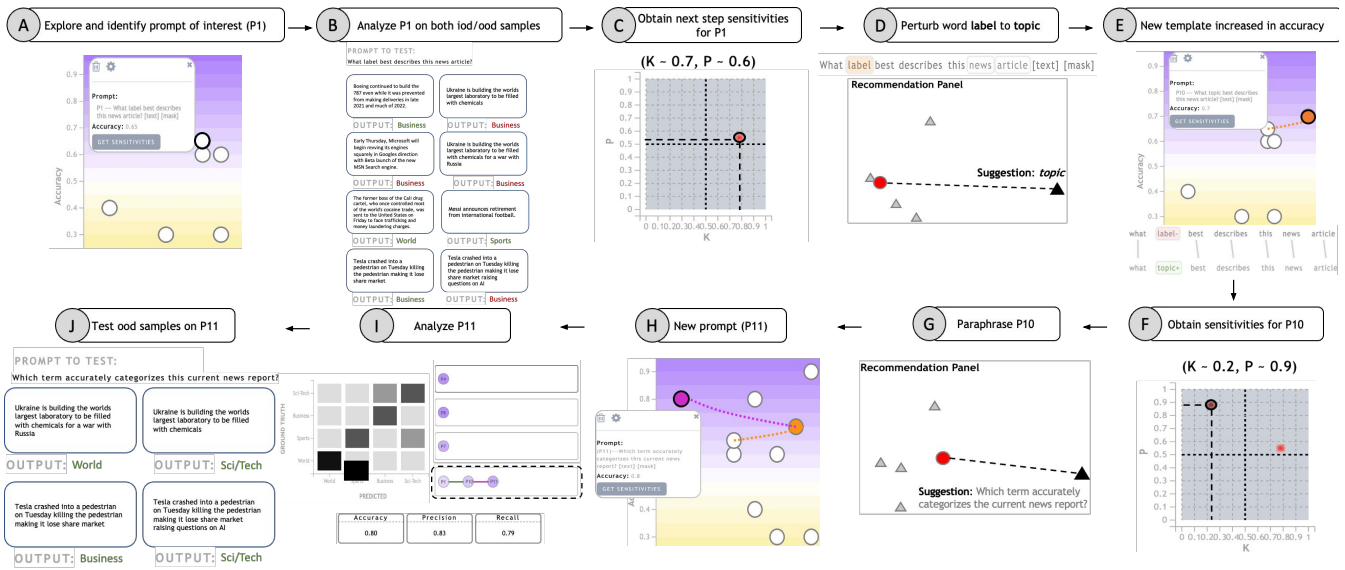}
    \vspace{-3mm}
   \caption{Case study \#1 using linguistic perturbations (keywords and paraphrasing) on the RoBERTa-base model for zero-shot settings. In a two step perturbation the accuracy of the prompt template increases from a 60\% accuracy to a 80\% accuracy on the test data set.}
   \label{fig:use_case_1}
\end{figure*}

\begin{figure*}[t]
   \centering
    \includegraphics[width=\linewidth]{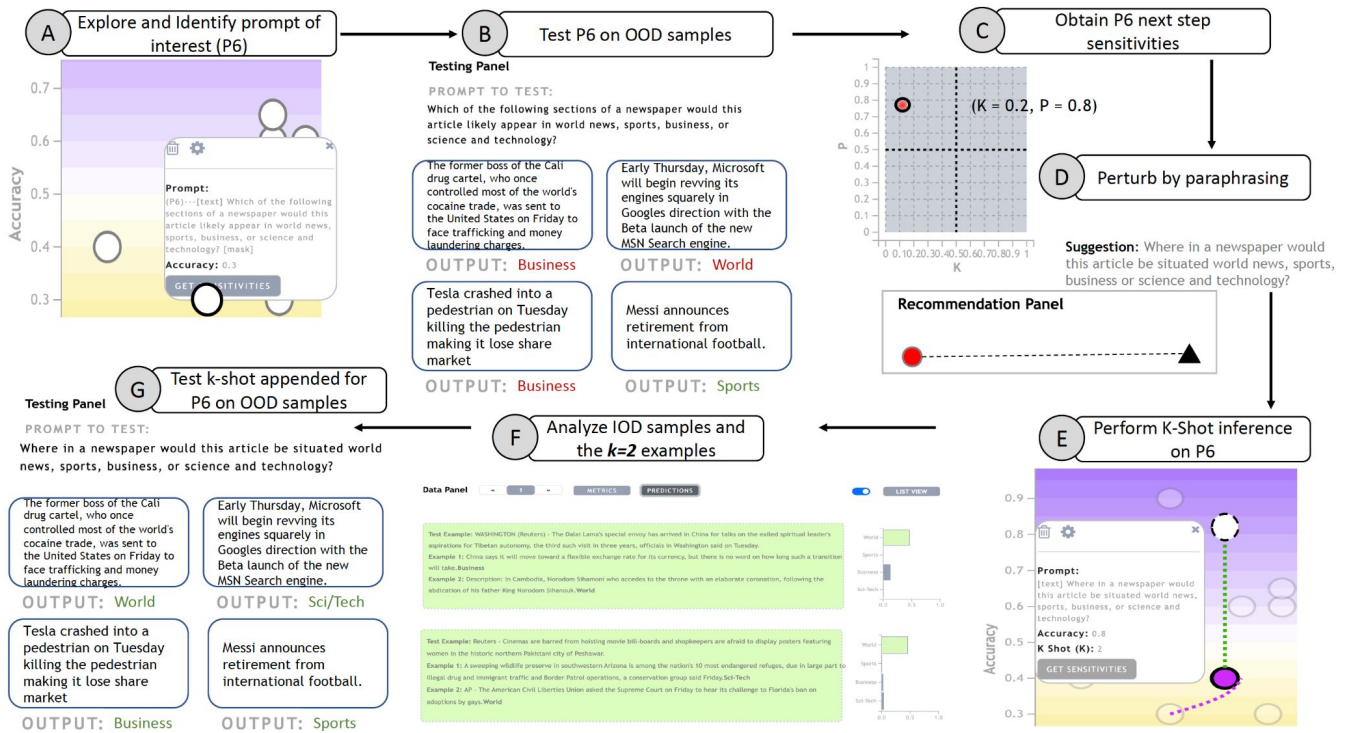}
    \vspace{-3mm}
   \caption{Case study \#2 using Contextual perturbations GPT-2 model for K-shot settings with optimal K returned as $k=2$. The accuracy of the prompt template increases from 30\% to 80\% by adding few-shot examples recommended by the system.}
   \label{fig:use_case_2}
\end{figure*}

\section{Frontend System Design : \textsc{\name{}}}
\label{sec:frontend}


In this section, we describe the \textsc{\name{}} interface, 
designed to support \textbf{(G1--G5)} by letting users iteratively explore, perturb, and test prompts and prompt templates.
Figure~\ref{fig:teaser} shows the interface, which is composed of six linked panels \textsf{(A)}--\textsf{(F)}.
Henceforth, we refer to the panels and sub-panels of Figure~\ref{fig:teaser} without pretending ``Figure'' to save repetition in this Section.

\textsf{(A)} The \textbf{Control Panel} lets users select a desired dataset and LLM~\textsf{(a1)}. A Prompt Editor~\textsf{(a2)} lets users enter personalized prompts, which on submission, loads the newly written prompt template on the prompt canvas panel. 
The Perturbation Sensitivity plot~\textsf{(a3)} is a scatter plot representing (x-axis) the average heuristic accuracy for next step perturbation on changing a keyword for a prompt, and (y-axis) the same for next step perturbation of paraphrasing the prompt template. 
This lets the user see what kind of linguistic perturbations to a prompt template can potentially increase its performance in the next step for 0-shot settings \textbf{(G4)}. Below, a [TEXT] toggle~\textsf{(a4)} highlights prompt templates in which the data point is appended at the beginning of the prompt template, to help users compare the performances of the templates based on their structure. 
The bottom of this panel provides a legend~\textsf{(a5)}. 

\textsf{(B)} The \textbf{Prompt Canvas Panel} provides an overview of the prompts being retrieved, written, and altered. It contains the following parts:
The primary chart (showing a purple-to-grey background) shows loaded prompts as circles~\textsf{(b1)}. The vertical position maps the accuracy of the prompt on the testing dataset, and the horizontal position arranges the circles as a 1-dimensional t-SNE projection, where more similar prompts are placed closer together, as this can help users understand if (and when) similarly worded prompts also have similar performance~\textbf{(G1)}. At right, a histogram shows the frequency of prompt templates based on performance~\textsf{(b9)}. 
 
Upon hovering, a tooltip is displayed~\textsf{(b2)}, and the prompt is shown in the panel's header~\textsf{(b3)}. The tooltip shows the template number, prompt template, its accuracy, and user controls, which allows the user to choose from three improvement/perturbation options: keyword suggestions, paraphrase suggestions and the addition of k-shot examples~\textsf{(b5, b8)}, along with a ``Get Sensitivities'' button~\textsf{(b6)} to calculate the heuristic perturbation accuracies for the next step, and a delete icon to erase the prompt~\textsf{(b7)}. 

When selecting an improvement option~\textsf{(b8)}, the prompt in the panel's header~\textsf{(b8)} is highlighted: The ``Suggest Keywords'' button highlights non-stopwords in the prompt template in the panel's header, and displays contextually similar words for the clicked word in the recommendation panel \textsf{(E)}. The ``Suggest Paraphrasing'' button likewise generates new paraphrases of the prompt template in the recommendation panel. The ``Get K-Shot Example'' button adds an optimal set of $k$ in-context examples to the prompt and displays the k-shot prompt template on the Prompt Canvas Panel. Clicking on a prompt template color-codes the data panel for instance-based analysis and populates the template to test in the Testing Panel \textsf{(F)}. 

The panel's footer~\textsf{(b10)} acts as a version comparison tool between two iterations of prompt templates: users can track words and position changes, and see what is added, removed, or maintained between between the two versions.

\textsf{(C)} The \textbf{Data Panel} supports detailed instance analysis for the selected dataset \textbf{(G2)}. Two data points are shown on a page; the user can navigate pages to see other points~\textsf{(c1)}. Correctly classified data points have green backgrounds (incorrect are red)~\textsf{(c2)}. Clicking the Predictions button~\textsf{(c4)}  toggles bar charts for each data point displaying the logits for each class~\textsf{(c5)}. In cases of correct classification, the green-striped bar represents the class to which the data point was classified. For incorrect classifications, the green stripe denotes the ground truth, while the red-striped bar represents the predicted class .

Clicking the Metrics button~\textsf{(c3)} shows accuracy metrics, including precision, recall, and a confusion matrix, for each prompt template. These metrics provide a global quantitative measure of the effectiveness of the prompt template, enabling users to evaluate its performance across the tested dataset.

When a k-shot prompt template is selected, this only displays one data point appended with the corresponding optimal k-shot examples and its logits. Other data points appended with their optimal k examples can be viewed by using the page navigator button.

\textsf{(D)} The \textbf{LeaderBoard and Provenance Panel} serves as a tracking mechanism to monitor the various versions of a prompt template. Each version is shown inside a rectangle band, ordered in descending order based on the heuristic accuracy of the prompt template. Inside each band, a circle labeled with the prompt serial number denotes the initial seed prompt template, color coded purple-to-yellow corresponding to the accuracy levels in the prompt canvas panel.

Hovering over a rectangle band~\textsf{(d1)} highlights the associated prompt template in the Prompt Canvas Panel. As the prompt template undergoes iterations in the prompt canvas panel, new linked circles are added to the right of the original seed prompt template; these symbolize the type of perturbation applied (indicated by the legend in the control panel). Clicking inside the rectangle band~\textsf{(d2)} populates the Prompt Canvas Panel's footer~\textsf{(b10)}, showing a textual comparison of differences between the various prompt template versions.

\textsf{(E)} The \textbf{Recommendation Panel} shows either keyword or paraphrase recommendations, based on the perturbation choice made in the Prompt Canvas Panel \textbf{(G3)}. A red circle that designates the word or prompt template currently undergoing modification; triangles represent suggested perturbations. Points are placed based on their similarity using a t-SNE layout, with contextually similar points positioned closer to the red dot than those that are farther apart. A hover tooltip~\textsf{(e1)} shows suggested keywords or paraphrases. Clicking a triangle~\textsf{(e2)} initiates the modification of the initial prompt template, with the newly altered prompt being loaded into the Prompt Canvas Panel. The modified prompt also has a link connecting it to the old prompt template, which is color-coded to indicate the type of perturbation applied.

\textsf{(F)} The \textbf{Testing Panel} supports testing a prompt template on a selection of desired data points, including both in-distribution or out-of-distribution (OOD) samples~\textbf{(G5)}, to validate the user's comprehension of the generated prompt template. Examples can be entered into a text box~\textsf{(f1)}; once submitted, the output is generated based on the predictions made by the LLM~\textsf{(f2)}.

\section{Case Studies}
\label{sec:case_studies}

To illustrate how \textsc{\name{}} can explore, perturb, and iterate prompts for higher accuracy, we present a usage scenario for zero-shot prompting for a topic classification task  on the ag$\textunderscore$news benchmark dataset. 
We adopt the perspective of Gary, a non-expert in AI/NLP, who has a set of OOD snippets of news articles he wishes to both classify and validate his own conclusion(s) about the task. These OOD samples are a set of recent news snippets obtained from the internet.
This use case is also presented in the demo video, found in the supplementary materials. 
To demonstrate \textsc{\name{}}'s ability to support K-shot prompting, we additionally include a second use case scenario (also for the topic classification task) in Appendix~\ref{sec:usecase_2}, due to page constraints.

\subsection{Use Case 1: Improving Zero-Shot Prompting}
\label{sec:use_case_1}
Gary's analysis and his specific actions are shown in Figure~\ref{fig:use_case_1}.
Gary selects the ag\textunderscore dataset and Roberta-base model, which populates the Prompt Canvas with ten seed prompt templates (P1-P10), each with a corresponding accuracy achieved by the prompt on a testing dataset, which Gary evaluates by hovering over each prompt. Upon closer inspection, he discovers that pairs of prompts differ only in the order in which the textual data, i.e where the [text] tag occurs in a prompt template, with one variant placing the textual data before the prompt and the other placing it after. The prompts with differing textual data orders appear at the same vertical position on the chart, indicating that their content is identical apart from the order of the textual data.

To investigate whether the order of textual data has an effect on prompt performance, Gary clicks the [TEXT] toggle in the Control Panel. He observes that prompts with textual data appended before the prompt tend to perform worse than those with textual data appended after the prompt. Gary uses this information to design his own custom prompt templates. Upon exploring, he discovers a simple and straightforward prompt template (P1: ``\textit{What label best describes this news article? [text]}'') which achieves an accuracy of 60\% on the testing dataset. However, upon examining the logits for each test data point and checking the data panel, Gary notices that the prompt template appears to be somewhat biased towards Business and Sports. 
The Business and Sports category is consistently predicted as the second most likely label after its first label for many test data points, even for those which are correctly predicted.

To further investigate this bias, Gary tests several examples from his own test set by clicking on the circle, which populates the prompt template on the testing panel. 
He enters eight OOD samples, some of which are ambiguous, to assess the model's performance. 
Figure \ref{fig:use_case_1}\textsf{(B)} displays the eight samples, where four are correctly classified. 
For example, the news snippet, ``\textit{Boeing continued to build the 787 even while it was prevented from making deliveries in late 2021 and much of 2022}'', which might appear to be a Sci/Tech news, is correctly classified as a Business news by prompt P1. 
However, some samples, such as ``\textit{Early Thursday, Microsoft will begin revving its engines squarely in Google's direction with the Beta launch of the new MSN Search engine.},'' appear to be Business news but are actually Sci/Tech news. 
Gary also enters news articles designed to confuse the LLM, such as ``\textit{Ukraine is building the world's largest laboratory to be filled with chemicals}.''

By examining the model's predictions for the aforementioned snippet, he aims to determine whether certain words (such as \textit{Ukraine} and \textit{world}) bias the LLM towards certain labels such as Sci/Tech or World news. He notices that the model incorrectly classifies the snippet as World news rather than Sci/Tech, as it contains words such as chemicals and laboratory, which are more typical of Sci/Tech news. 
Similar to the in-distribution samples, Gary observes that most news articles are being predicted as Business or Sports, and even correctly classified articles tend to belong to these two classes.

\textsf{(C)} Gary then clicks on the ``Get Sensitivities'' button for P1 to get an idea of the perturbation type to alter the prompt template. 
A red dot is loaded in the Sensitivity Panel with a 0.6 value for nex step paraphrasing and a 0.7 average accuracy for next step keyword-based change. 
He then invokes the ``Suggest Keywords'' action. 
\textsf{(D)} This draws a bounding box around three words in the prompt template: \textit{label}, \textit{news}, and \textit{article}. 
He first clicks on \textit{news}, which does not yield any recommendations except the word itself, and then clicks on \textit{label}. 
This populates the recommendation panel with the suggestions, ``criteria, tag, name and topic.'' 
Gary selects ``topic,'' as this word has the maximum distance from the red dot while remaining relevant to the topic classification task of ``label.''

\textsf{(E)} Clicking on the triangle populates the Prompt Canvas Panel with a new, altered prompt. 
This accuracy of the new prompt (P10) has increased to about 70\% on the testing set. 
He again checks the previously mentioned eight data points using the testing panel and finds that out of the eight articles, five are correctly classified or classified into classes that match his mental model of what those snippets should be categorized into. 
However, one snippet (``\textit{Tesla crashed into a pedestrian on Tuesday killing the pedestrian making it lose share market now raising questions on AI.}'') is still being classified as Business, and the prompt template is unable to classify it as Sci/Tech news. 

He reviews the LeaderBoard Panel and notices that the original template (P1) is now linked to (P10) in a way that indicates the prompt template's performance has increased. Clicking on the rectangle band shows the version control between the two templates, and Gary sees that only one word has been altered (\textit{label} is now \textit{topic}) with every other word being the same. 

To ensure that the results of the data points he is entering are consistent, Gary tries to further improve upon the prompt (P10). 
\textsf{(F)} He again clicks on the ``Get Sensitivities'' button, now for (P10), loading its red dot in the Sensitivity Panel with a higher next step accuracy for Paraphrase based perturbation (0.9) and a very low (0.2) Keyword based perturbation. 
\textsf{(G)} Gary clicks the ``Suggest Paraphrases'' option for (P10), which populates the Recommendation Panel with five paraphrased recommendations. 
Most of the recommendations seem to align with the task at hand and are also different from one another: 
``\textit{Tell me the best topic for this news article?},'' ``\textit{What category would this news article best be in?},'' and ``\textit{Which term accurately categorizes this current news report?}'', which Gary chooses.

\textsf{(H)} Clicking this recommended paraphrase loads a new prompt (P11) in the Prompt Canvas Panel, with a purple-colored link to highlight that this is a paraphrase-based change, with the accuracy increasing to 80\%. 
Clicking on the Data Panel showcases its correct and incorrect data points.
\textsf{(I)} The confusion matrix also is darker at the diagonals suggesting that the ground truths are equal to the predictions, and the precision and recall of the template have also increased. This new prompt is also now placed at the fourth position among all the prompts on the LeaderBoard.

\textsf{(J)} Gary finally tests all eight data points from \textsf{(B)}, which were meant to check the robustness of the prompt generated and he finds that all the data points have been correctly classified. This helps Gary gain confidence in the prompt template he has generated by keywords and paraphrase-based alterations for zero-shot settings.

\subsection{Use Case 2: Adding Few-Shot Priming Examples}
\label{sec:usecase_2}

In contrast to Case Study 1, Gary now seeks to explore the impact of in-context examples on the performance of the generative model GPT-2. Figure~\ref{fig:use_case_2} shows his actions.

He selects the same ag$\textunderscore$ news dataset but switches the model to GPT-2. \textsf{(A)} He observes that most prompt templates yield accuracies higher than 50\%, except for a single prompt template (P6: ``\textit{Which of the following sections of a newspaper would this article likely appear in world news, sports, business, or science and technology?}''), which records the lowest accuracy score of 30\% among all templates. Upon selecting (P6), Gary realizes that most of its predictions are incorrect. \textsf{(B)} He further evaluates the prompt template using the Testing Panel, employing OOD examples such as ``\textit{The former boss of the Cali drug cartel, who once controlled most of the world's cocaine trade, was sent to the United States on Friday to face trafficking and money laundering charges},'' ``\textit{Tommy Fury handed the YouTuber-turned-boxer the first loss of his career on Sunday night at Diriyah Arena in Riyadh, Saudi Arabia},'' and ``\textit{Reuters - Cinemas are barred from hoisting movie bill-boards, and shopkeepers are afraid to display posters featuring women in the historic northern Pakistani city of Peshawar.}'' However, none of the answers generated match Gary's expectations as seen in Figure \ref{sec:usecase_2}.

\textsf{(C)} Clicking on the ``Get Sensitivities'' button for (P6), Gary notices that the paraphrase-based perturbation exhibits an average accuracy of 20\%, whereas the keyword perturbation-based accuracy performs even worse with a 10\% accuracy score. Although the perturbed accuracies are lower, they are averaged over numerous perturbed samples, prompting Gary to opt for paraphrasing the prompt template to enhance the performance of (P6). \textsf{(D)} He clicks on ``Suggest Paraphrases,'' which provides him with a single paraphrased suggestion: ``\textit{Where in a newspaper would this article be situated: world news, sports, business, or science and technology?}'' Gary deems the paraphrase appropriate and selects it, creating a new prompt template in the Prompt Canvas panel. The performance of this perturbed prompt subsequently increases to 50\%.

\textsf{(E)} Gary decides to augment the test dataset with k-shot- be consistent, here and throughout the paper examples to investigate if the prompt template performs better with additional examples. He clicks on the ``Add K-Shot Examples'' button, and the same prompt template appears in a green dashed border, performing exceptionally well with an accuracy score of approximately 80\%. Gary selects the k-shot prompt template and evaluates the data points added to the test dataset. \textsf{(F)} He observes that the logits for the correct predictions heavily lean towards the correct predictions, with significantly lower logits for other classes. This insight leads Gary to understand that GPT-2 becomes biased as more examples are added, with the logits strengthening towards the correct class. \textsf{(G)} He finally evaluates the prompt template using his own examples and notes that the LLM generates a correct output for all the examples, as seen in Figure \ref{fig:use_case_2}.

\section{Evaluation}
\label{sec:evaluation}

To empirically evaluate \textsc{\name{}}, we conducted a controlled, within-subject user study. 
We recruited ten participants (u1--u10) who, notably, were not experts in NLP, or more specifically LLMs, except for occasional usage of ChatGPT (e.g., as a replacement for Googling). 
This evaluation serves two purposes: (1) To understand if and how \textsc{\name{}} aids users to iteratively design prompt templates based on linguistic and contextual system recommendations; and (2) to compare \textsc{\name{}}'s visualization and recommendation approach against a baseline interface lacking these features.

\begin{figure}[t]
   \centering    \includegraphics[width=\columnwidth]{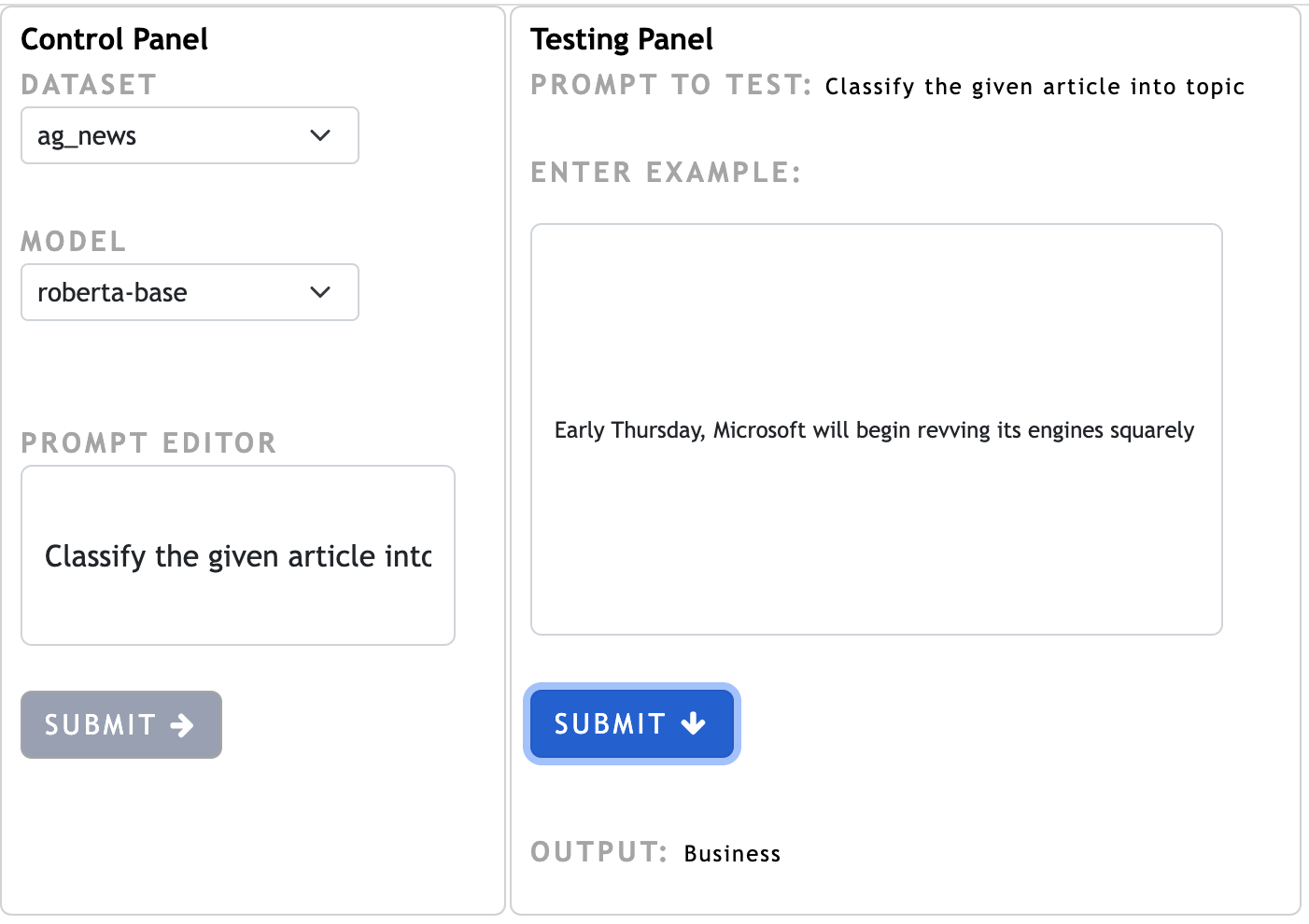}
   \caption{The baseline interface for the user study inline to most commonly available interfaces available for prompting LLMs.}
   \label{fig:baseline_interface}
\end{figure}

\subsection{Study Design and Setup}

\textbf{Baseline Interface.} As a baseline for comparing against \textsc{\name{}}, we designed a straightforward interface (see~Figure~\ref{fig:baseline_interface}) featuring an input area for prompt templates and a box for users to input examples. 
This interface is typical of prompting interfaces commonly available on the internet such as ChatGPT and GPT-3 which are already used by non-expert users. 
To ensure a fair comparison, identical datasets, and LLMs were utilized for \textsc{\name{}} and the baseline.


\textbf{Domains and Models.} One domain (i.e., ag$\textunderscore$news) and two language models (i.e., RoBERTa-base and GPT-2) were used. 
We perform the keyword and paraphrase alteration based task on RoBERTa-base for zero-shot settings, and the k-shot inference task using GPT-2, doing topic classification. 


\textbf{Design.} The study consisted of five stages:

\textit{(1) Interface Assignment and Training Stage.} The participant was assigned an initial interface. A hands-on training was given to explain system features and interactions. Participants asked as many questions as they want and were given a chance to play with the interface until they are ready to proceed.

\textit{(2) Task Stage.} Participants were asked to perform the following two tasks: (t1) linguistic alterations (using keywords and paraphrasing), and (t2) contextual few-shot inferencing, to improve the performance of a given prompt template.
Participants were able to make as many perturbations as they wanted using the functionality and features of the assigned interface until they were satisfied with the prompt template. 

No time limit was set for this stage, for two reasons: (1) We wanted to measure the confidence that non-experts had in their ``completed'' prompt template, rather than a non-final version restricted by time; and (2) we also wanted participants to gain sufficient usage to allow them to subsequently assess each interface in terms of required cognitive effort.


To finish this Stage, participants completed a short survey by rating the tracking abilities and the cognitive efforts required to change the prompt template, based on a 7-level Likert scale.

\textit{(3,4) Repeat Training and Task Stages with the other Interface.} Participants switched to the other interface and repeated the Training and Task Stages. Trials in the second iteration of the task utilize a second prompt template. To minimize potential confounds, the order of interface assignments, the selection of prompt templates, and the trial order were counterbalanced among participants.

\textit{(5) Freeform Analysis Stage}. Participants were allowed to freely use and explore \textsc{\name} for any of the datasets or models implemented.  No specific task was assigned, but participants were encouraged to put themselves into the following motivating scenario: \textit{They were given a set of ambiguous data points and were not sure in which class the data points would be classified. 
They had to come up with prompt templates which could lead to high accuracy.} 

In this Stage, we primarily wanted to assess the overall usability of \textsc{\name} (e.g., its general user experience and specific interface features), thus the baseline interface was not used. Participants had ten minutes to complete this stage and used a think-aloud protocol to verbalize their cognitive processes. 
At the end of the Stage, participants completed a short usability survey, and if desired, they were encouraged to provide additional comments about \textsc{\name} and the baseline. To account for the complexity and novelty of the interfaces and study tasks, an administrator sat beside the participants, to answer questions or help them if they became stuck or confused.

\textbf{Participants and Apparatus. }Ten graduate computer science students were recruited from \textsf{<Anonymous University>} (average age
= 24.6, SD = 1.42; 6 males, 4 females). Though some of the graduate students were familiar with general AI/ML concepts, all reported little-to-no experience in NLP and LLMs. Each session lasted 30–45 minutes, completed using Google Chrome in full screen mode at $3840\times2160$ resolution. The study was completed in a quiet, office-like environment with no distractions.

\subsection{Study Results}

\subsubsection{Task Stage Performance}
In the Task Stage, we primarily report on collected survey ratings from the participants about the cognitive effort required, tracking abilities and confidence of their acquired prompt template for both interfaces. Where applicable, we report Mann-Whitney U tests to indicate if there is a statistical difference between \textsc{\name} and baseline (using a threshold of $p=0.05$) in terms of ease of generating good performing prompts by providing $U$ and $p$ values.

\begin{figure}[t]
   \centering
    \includegraphics[width=1\columnwidth]{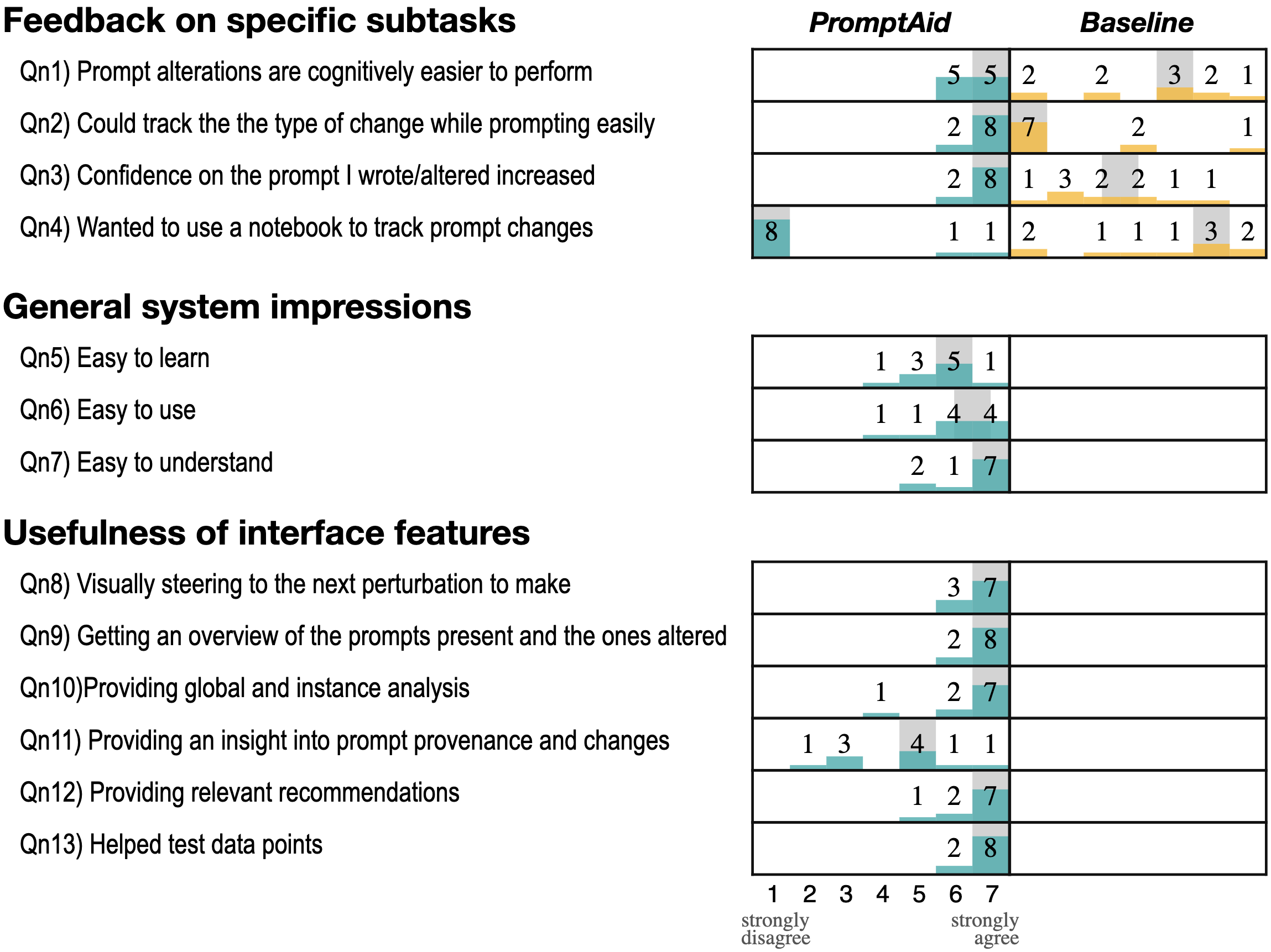}
    \vspace{-3mm}
   \caption{Participant ratings from the user study; median ratings are indicated in gray.}
   \vspace{-4mm}
   \label{fig:study1_ratings}
\end{figure}

Four questions were asked to participants about the cognitive effort required, knowledge about the change being made, need to track prompt template changes, and confidence in the prompt templates they reach, shown in Figure \ref{fig:study1_ratings} (Qn1--Qn4). 
For each question, \textsc{\name} performed significantly better in terms of the cognitive effort required while prompting $(U = 12.5, p < 0.005)$, tracking the kind of change being made $(U = 6, p < 0.005)$, ease of tracking $(U = 23.5, p < 0.05)$ and confidence acquired in the prompt template reached $(U = 1, p < 0.005)$. 
These results indicate that \textsc{\name} was not only successful in aiding prompt changes but also helped users track, compare and analyze prompt templates over iterations. 


\subsubsection{Freeform Stage: User Comments and Survey Ratings}


We next report comments and feedback collected during and after the Freeform Analysis Stage. Figure~\ref{fig:study1_ratings}(Qn5–-Qn13) shows survey feedback about the system during this stage. \textsc{\name}'s functionality and interface feature were highly rated by almost all the participants (as the baseline was not used in this stage, it does not have corresponding ratings for these questions). 
We performed an open coding on participant verbalizations (think aloud and additional commentary), and discuss both positive feedback as well as some suggested system improvements below, in the context of the \textsc{\name{}}'s design goals \textbf{(G1–-G5)}.

\textbf{(G3, G4) Visual steering and recommendations were preferred over baseline.} All ten participants preferred \textsc{\name{}}'s visual aids during prompting (compared to the baseline), and were able to obtain better-performing prompt templates over iterations with less cognitive effort. Several participant comments emphasized this:
``\textit{The visualization panels are basically pruning my search in the prompt space. I know what change to make in the next step to make my prompt template better}'' (u7). ``\textit{The recommendation panel was not only helpful in giving me new words or paraphrases, but it actually was helping me think newer words which I normally wouldn't think}'' (u4). 

Participants also explicitly described that the baseline required more cognitive processing: ``\textit{I could think of one word for the first change but then it gets harder to think of more changes to the template over steps}'' (u10). ``\textit{Prompting in baseline was harder as after a point I couldn't think of more changes but the interface was still giving me new suggestions for the same prompt template}'' (u9). ``\textit{Thinking in baseline to come up with synonyms was still fine but paraphrasing was hard. It was a bit uncomfortable}'' (u2).
These comments echo the Likert scale ratings in Figure~\ref{fig:study1_ratings}.

\textbf{(G1, G2, G3, G4, G5) Prompt template improvement and tracking across panels.}
All of \textsc{\name{}}'s six linked panels were used to contextualize prompt template performances and to iteratively validate results. Five participants (u1, u2, u3, u6 ,u8) mentioned that the Prompt Canvas Panel was extremely useful for a global view of prompt templates, comparison purposes, and keeping track of the changes they were making. 
``\textit{In baseline I couldn't make sure if my prompt was doing well on just one point or globally on a bunch of other points, Prompt canvas was really useful}'' (u2). ``\textit{Prompt canvas was really useful and helpful to track the changes I made to the template, it sort of acted like a global view unlike baseline where I didn't know how well my prompt was performing}'' (u1).

Four participants (u4, u5, u7, u9) especially liked the Perturbation Sensitivity Panel. ``\textit{The sensitivity panel was acting like a direction to let me know which way or change I need to make. I didn't have to think of it, unlike baseline}'' (u7). These four participants also found the recommendation panel to be the most useful: ``\textit{The recommendation plot was useful for me. I could think of newer words and also in my perturbations I found words or paraphrases farther from the red dot perform way better than recommendations near the red dot}'' (u6). ``\textit{Since I did the visual interface first I knew which word to change for the prompt template in the baseline interface but if I hadn't known this it would have taken me longer to think of this for the baseline}'' (u5). 

Almost all participants also mentioned that the Data Panel, along with prediction bars and confusion matrix, was useful for analyzing prompt templates. ``\textit{I used the Data Panel to see if the logits were biased to a certain class or not and used it to compare new iterations of the same prompt.}'' (u6). ``\textit{The confusion matrix is very useful, all I need to keep in mind is make the diagonals the darkest in color}'' (u8). 

All ten participants also described that the Testing Panel, while simple, was especially useful to validate their prompt templates as iterated. ``\textit{This is similar to baseline, but testing data points along with all these other panels is more useful than just blindly testing like I was doing in the base interface, and now I am more sure of my prompt when I test}'' (u8).


\textbf{(G3) More user controls for in-context examples for experimentation purposes.} All ten participants found the in-context example recommendations useful to improve the performance of the prompt template, and many commented that it increased their creativity: ``\textit{I never knew adding examples can increase the accuracy of the same prompt by so much, and the fact that I don't have to think of those examples is very convenient}'' (u10). ``\textit{I trust the backend with these optimal K values and the examples appended to the test data point, I would have never thought of those on my own.}''(u4).

One suggestion, made by four participants (u1, u2, u8, u10), was to have a panel that allowed users to pick the k-shot examples themselves from a larger set to determine the best-performing prompt. ``\textit{I actually trust that the system has come up with good examples for the k-shot setting, the accuracy here increased by 40\% but it wouldn't hurt to have an option where I too can play with examples I want to enter instead of just accepting the suggestions}'' (u10). ``\textit{I like the k-shot example suggestions and would definitely use it. It performs the best among all the changes I made to the prompt. But if I enter fake news to see if the prompt template does well or not I would want an option to enter fake examples for them as well just to play around}'' (u8). We considered the approach during \textsc{\name}'s prototyping, refer to section \ref{sec:key_para_kshot}, but it was omitted as we wanted to not overwhelm non-technical users. 
However, even without controls to choose their own examples, participants were satisfied with the examples recommended for the test data point.

\textbf{Peaking inside the LLM to contextualize behavior.} 
One suggestion, made by three participants (u2, u3, u8) was to provide explainability in terms of the saliency of the text being entered into the LLM, to understand what words were focused on more during predictions. ``\textit{I wish there was a panel that showed words which were more important for the [LLM] while giving out a prediction. The outputs match my expectations, but I would want to see something like text highlights as well}'' (u2). ``\textit{Knowing why the model works on one prompt and doesn't on another is something I would like to see}'' (u8).
While we agree that a gradient-based saliency would aid users, computing such saliency for LLMs is computationally prohibitive, see Section~\ref{sec:discussion}.

\textbf{On-demand training to improve usability.} Overall, participants found \textsc{\name} easy to learn, use, and understand (e.g., see Figure~\ref{fig:study1_ratings}(Qn5--Qn7)), though two participants (u3, u8) mentioned that additional training time would help them more intuitively understand the system’s functions and improve the overall user experience. Each suggested providing on-demand user guides, tutorials, or breadcrumbs within the interface. ``\textit{I really like the visual interface but before you explained it to me I was a little overwhelmed, so some sort of a manual or read me file would be really useful}'' (u3). ``\textit{The interface was easy to use after you explained it to me, but if some sort of pop-ups could be made in the interface to tell the meaning of those panels it would really help us to use it on our own.}'' As previously mentioned, this functionality was not implemented during the study as the administrator could assist participants who were stuck or confused. 
\section{Discussion}
\label{sec:discussion}
We view \textsc{\name} as a first attempt to make a generalizable visual interface to support an iterative exploration of prompt space, augmented with AI-backed recommendations for novice users. Here, we discuss takeaways and lessons learned from our development and evaluation of \textsc{\name}, such as how visualization-based approaches can make prompting easier and with a lower cognitive load for non-experts, as well as some current limitations in \textsc{\name} that can be addressed in future efforts.

\textbf{Prompting is still hard for non-expert users, but visual interfaces can significantly help.} Despite prompting being hailed as a method to democratize machine learning for the public, our pre- and post-studies with NLP experts and non-expert users suggest that there are still barriers to be overcome, not only in terms of cognitive effort required but also in terms of ease of prompting and the domain knowledge required for the task at hand.

To address these barriers, we proposed \textsc{\name}, a multi-panel visualization tool that was found to be highly useful by our non-expert study participants. The system effectively enabled users to experiment with prompts and make context-specific perturbations much more easily than a comparable baseline interface. Several participants noted that \textsc{\name}'s visual interface not only helped them cognitively when changing a prompt, and achieving better-performing prompts, but it also worked as a creativity tool, that helped them think of new words or new ways of phrasing statements that they normally would not have thought of on their own. We believe that \textsc{\name} can also serve as a stepping stone for researchers to identify further pain points faced by users and to build more accessible systems in the future.

\textbf{Research directions from our experiments. }While performing initial experiments in the backend on keywords, paraphrases, and k-shot-based perturbations we found that the addition of k-shot examples on average increased the prompt template's accuracy by $\approx$ 35\%, paraphrasing by $\approx$ 20\% and keywords by  $\approx$ 10\%. (We obtained these results by averaging the performances of prompt template perturbations three different times.) We also found that adding the test data point at the end of the prompt templates had a higher accuracy than the test data points appended at the beginning of the prompt templates. To understand why a certain type of alteration performs better than others, we need more research not only in terms of empirical experiments but also in terms of greying the black box to understand an LLM's behavior. Results from such analyses can be leveraged to make better design decisions in prompting. 

Due to the recency of prompt engineering, we tested \textsc{\name} on two currently-important strategies for perturbation (linguistic and contextual). However, there are many other nuanced (and likely interconnected) factors that can influence an LLM's outputs, such as prompt length, structure, and the type of language model used~\cite{jiang2022promptmaker, liu2023pre}. Prior work~\cite{liu2023pre} has shown that prefix prompts tend to work better on generative models and cloze-style prompts have a better performance on masked language models. While these are important factors that play an important role in designing better prompts, such in-depth analysis of LLMs itself calls for more extensive empirical experiments, both within the NLP community, but also in developing novel interactive and visual tools, in the future.

\textbf{Integrating saliency-based methods to promote trust in LLMs.} One current constraint in \textsc{\name} is that we do not ``open the black box'' to promote explainability and trust for users. As AI and NLP models (including LLMs) become increasingly prevalent in our everyday lives, it is essential to make these models more interpretable not only for AI/NLP experts but also for non-expert users~\cite{gdpr}. As an example, several study participants expressed the desire to see which words the LLM focused on when generating output. While there have been efforts to visualize feature attributions and neuron activation patterns for LLMs, such methods can be extremely computationally expensive and time-consuming~\cite{alammar-2021-ecco}. In the future, we intend to investigate more transparent interaction paradigms, to support non-experts in gaining insight into the decision-making process of the LLMs, without incurring significant overhead or requiring in-depth technical expertise.

\textbf{Supporting prompt provenance.} Another area for investigation pertains to developing enhanced interfaces for monitoring prompt provenance across iterations. While \textsc{\name} affords users the ability to monitor prompt provenance, creating more robust frameworks that can demonstrate changes based on individual words, trace parts-of-speech tags over various prompt iterations, monitor prompt performance over iterations, and compare performance across multiple LLMs could provide a more comprehensive understanding of which prompting techniques are more effective with specific LLMs. These frameworks may also assist domain experts in identifying sources of bias and errors in an LLM's output. Our future work aims to develop more intricate prompt-tracking interfaces.

\section{Conclusion}

\textsc{\name} is a visual analytic system that lets a human explore, perturb, test, and iterate over prompts to prompt a language model better. \textsc{\name} supports both masked and generative language models and is task agnostic. \textsc{\name} supports three types of context-specific changes to a prompt: keywords, paraphrasing, and few-shot priming examples. Results from a controlled user study found these visual encodings preferable to widely available state-of-the-art interfaces for prompting language models. Additionally, users found the visual steering in the interface to reach better prompts in a latent space very useful. Future work intents to expand on the other factors which can affect a prompt performance such as the length of the prompt, the structure of prompts, and providing users with more interpretability while prompting LLMs. 

\section{Appendix}

\subsection{Usage of a Constrained Testing Set}
\label{sec:usage_of_smaller_testing_set}

As described in Section~\ref{sec:dataset_and_model}, we test \textsc{\name{}} on a sample of 20 data points. Testing on a dataset of this size was deemed appropriate and sufficient for several reasons. First, current open-source LLMs tend to be slow, and testing a prompt on a smaller scale evaluates the prompt template faster, which significantly aids in faster exploratory analysis. Secondly, storing data for prompt templates in a database would not account for novel prompt templates that users enter, causing delays in backend processing and visual feedback. Furthermore, pre-calculating metrics and recommendations for perturbed prompt templates causes the prompt space to grow exponentially, which would make the system inaccessible, as people would only be able to enter prompts that are pre-stored and calculated in databases.

To address these concerns, \textsc{\name{}} provides a Testing Panel, where users can enter their own examples (which might not be included in the dataset itself). The frontend system is agnostic to the size of data points it is tested upon, allowing users to easily test their prompts on more samples if needed. 

\subsection{Use Case 2: Adding Few-Shot Priming Examples}
\label{sec:usecase_2}

In contrast to Case Study 1, Gary now seeks to explore the impact of in-context examples on the performance of the generative model GPT-2. Figure~\ref{fig:use_case_2} in the Appendix shows his actions.

He selects the same ag$\textunderscore$ news dataset but switches the model to GPT-2. \textsf{(A)} He observes that most prompt templates yield accuracies higher than 50\%, except for a single prompt template (P6: ``\textit{Which of the following sections of a newspaper would this article likely appear in world news, sports, business, or science and technology?}''), which records the lowest accuracy score of 30\% among all templates. Upon selecting (P6), Gary realizes that most of its predictions are incorrect. \textsf{(B)} He further evaluates the prompt template using the Testing Panel, employing OOD examples such as ``\textit{The former boss of the Cali drug cartel, who once controlled most of the world's cocaine trade, was sent to the United States on Friday to face trafficking and money laundering charges},'' ``\textit{Tommy Fury handed the YouTuber-turned-boxer the first loss of his career on Sunday night at Diriyah Arena in Riyadh, Saudi Arabia},'' and ``\textit{Reuters - Cinemas are barred from hoisting movie bill-boards, and shopkeepers are afraid to display posters featuring women in the historic northern Pakistani city of Peshawar.}'' However, none of the answers generated match Gary's expectations as seen in Figure \ref{sec:usecase_2}.

\textsf{(C)} Clicking on the ``Get Sensitivities'' button for (P6), Gary notices that the paraphrase-based perturbation exhibits an average accuracy of 20\%, whereas the keyword perturbation-based accuracy performs even worse with a 10\% accuracy score. Although the perturbed accuracies are lower, they are averaged over numerous perturbed samples, prompting Gary to opt for paraphrasing the prompt template to enhance the performance of (P6). \textsf{(D)} He clicks on ``Suggest Paraphrases,'' which provides him with a single paraphrased suggestion: ``\textit{Where in a newspaper would this article be situated: world news, sports, business, or science and technology?}'' Gary deems the paraphrase appropriate and selects it, creating a new prompt template in the Prompt Canvas panel. The performance of this perturbed prompt subsequently increases to 50\%.

\textsf{(E)} Gary decides to augment the test dataset with k-shot- be consistent, here and throughout the paper examples to investigate if the prompt template performs better with additional examples. He clicks on the ``Add K-Shot Examples'' button, and the same prompt template appears in a green dashed border, performing exceptionally well with an accuracy score of approximately 80\%. Gary selects the k-shot prompt template and evaluates the data points added to the test dataset. \textsf{(F)} He observes that the logits for the correct predictions heavily lean towards the correct predictions, with significantly lower logits for other classes. This insight leads Gary to understand that GPT-2 becomes biased as more examples are added, with the logits strengthening towards the correct class. \textsf{(G)} He finally evaluates the prompt template using his own examples and notes that the LLM generates a correct output for all the examples, as seen in Figure \ref{fig:use_case_2}.


\bibliographystyle{abbrv-doi}

\bibliography{template.bib}
\end{document}